\documentclass[newversion]{aa}
\usepackage{graphicx}

\begin{document}


   \title{Solar prominences with Na and Mg emissions \\
and centrally reversed Balmer lines}

   \author{G. Stellmacher\inst{1}
          \and E. Wiehr\inst{2}}

   \offprints{E. Wiehr}

   \mail{ewiehr@astrophysik.uni-goettingen.de}

   \institute{Institute d'Astrophysique (IAP), 98 bis Blvd. d'Arago, 
              75014 Paris, France
              \and
              Institut f\" ur Astrophysik der Universit\"at,
              Friedrich-Hund-Platz 1, 37077 G\"ottingen, Germany}

   \date{Received June 29, 2004; accepted Oct. 14, 2004}

\abstract
{}
{We study spectral lines in exceptionally bright solar limb prominences with pronounced 
sodium and magnesium emission and central reversion of the first two hydrogen Balmer lines.}
{We simultaneously measure the line profiles of H$\alpha$, H$\beta$, HeD$_3$, 
He\,II\,4685, He\,I\,5015 (singlet), NaD$_2$ and Mgb$_2$ using the THEMIS telescope 
on Tenerife.}
{We find that most prominences with significant NaD$_2$ and Mgb$_2$ emission show 
centrally reversed profiles of H$\alpha$ and occasionally even of H$\beta$. 
The strongest emissions reach integrated intensities $E\beta > 16 \cdot 10^4$ 
erg/(s\,cm$^2$\,ster). The centrally reversed profiles are well reproduced by 
semi-infinite models. The source function reaches $S_{\alpha} \le 36 \cdot 10^4$ 
erg/(s\,cm$^2$\,ster\,\AA{}) corresponding to an excitation temperature of
$T_{ex}\approx3950$\,K; here, the optical thickness of H$_{\alpha}$ amounts to 
$\tau\approx10.0$. The narrow widths of the NaD$_2$ and Mgb$_2$ profiles yield 
a non-thermal broadening of $V_{tu} = 5$\,km/s.} 
{} 

\maketitle

%

\section{Introduction}

The simultaneous occurrence of resonance lines with low ionization potential, like 
Mgb$_2$ and NaD$_2$, and of hydrogen and helium lines with much higher excitation and 
ionization energy illustrates the large deviation from LTE in atmospheres of 
solar prominences. Only few comprehensive sets of simultaneous emission data 
have so far been published, e.g., Yakovkin \& Zel'dina (1964), Kim (1987). High 
precision photometry of prominence spectra show for faint emissions 
($E\beta < 1\cdot 10^4$ erg/(s\,cm$^2$\,ster); corresponding to $\tau\alpha < 1.0$) 
an empirical relation between H$\alpha$ and H$\beta$ which is independent of the 
individual prominence (Stellmacher \& Wiehr 1994b). For brighter emissions, 
however, this relation depends on the prominence atmosphere. 
The present data are considered as an extension toward strongest emissions 
with $E\beta > 4\cdot 10^4$ erg/(s cm$^2$ ster). Such bright prominences are known 
to be cool, dense, and rather unstructured (Stellmacher \& Wiehr 1995). They allow 
a determination of upper limits of the H$\alpha$ source function as well as a 
quantitative analysis of the centrally reversed H$\alpha$ profiles and their 
representation by models.

%
%

\section{Observations and data reduction}

We observed with the French-Italian solar telescope THEMIS simultaneously the emission lines 
H$\alpha$, H$\beta$, HeD$_3$, He\,II\,4685, He\,I\,5015\AA{} (singlet) NaD$_2$ and Mgb$_2$ 
on October 18 and 23, 2000. For the two data sets the entrance slit of 0.5\,arcsec and 
0.75\,arcsec width, respectively, was aligned along the direction of atmospheric refraction
(i.e. perpendicular to the horizon). Exposure times of a few seconds gave about 2000\,counts 
for the brightest H$\alpha$ and 300 counts for the faintest Mgb$_2$ emissions.

The CCD-images were corrected for the dark and the gain matrices; the underlying stray-light 
aureole was subtracted using spectra from locations adjacent to the corresponding prominence. 
For the calibration of the prominence emissions, we took disk-center spectra and use the 
absolute intensities given by Labs \& Neckel (1970) in units of [$10^6$ erg/(s\,cm$^2$\,ster\,\AA)]:
I\,(6563\AA) = 2.86, I\,(5890\AA) = 3.34, I\,(5876\AA) = 3.36, I\,(5173\AA) = 3.93, 
I\,(50I5\AA) = 4.06, I\,(4861\AA) = 4.16 and I\,(4685\AA) = 4.3.

We determine the total emission $E$ ('line radiance') as the intensity integrated over 
the whole profile of the line: $E = \int I d\lambda$ [erg/(s cm$^2$ ster)] and give $I$ and $E$ 
in units of $10^4$ to enable an easy comparison with former data. H$\alpha$ solar survey 
images of the prominences under study (Fig.\,1) show that these low-latitude objects 
$(\vartheta < 32°)$ occurred under various aspect angles between 'face-on' and 'end-on'. 
An example of simultaneously observed CCD spectra is displayed in Fig.\,2.
                       
%
   \begin{figure}[ht]     
   \hspace{4mm}\includegraphics[width=8.0cm]{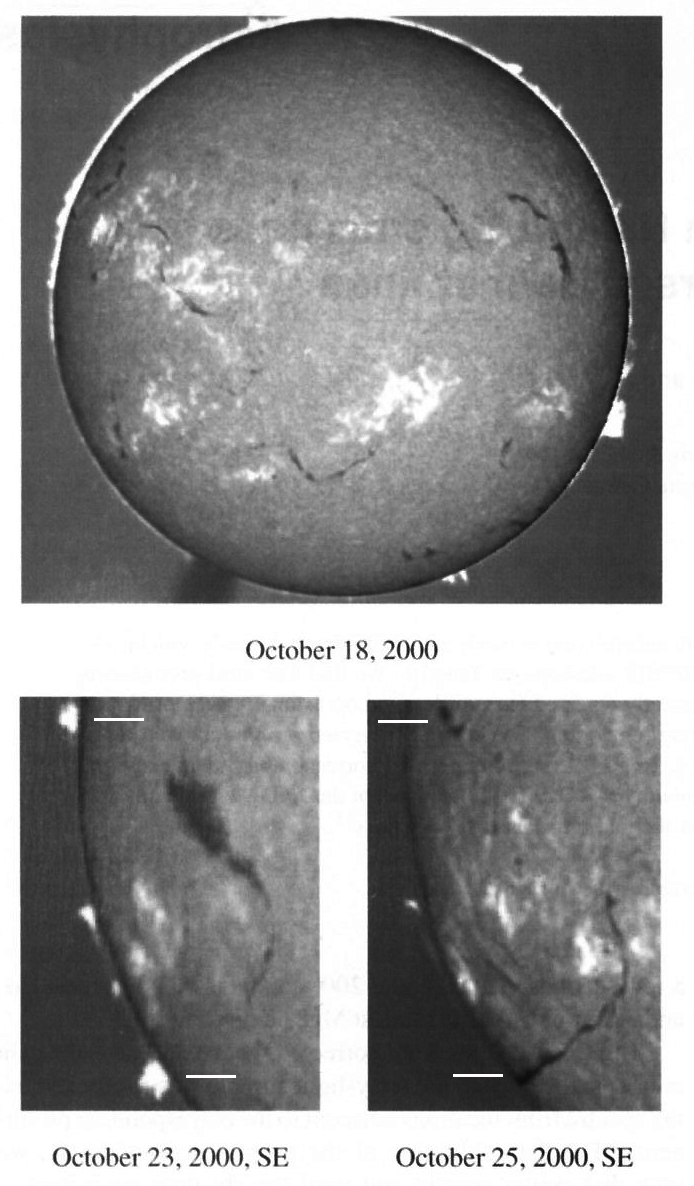}
   \caption{H$\alpha$ solar survey images (Meudon observatory) with the 
prominences at E/22$^o$N and W/24$^o$S observed on Oct.\,18 ({\it upper panel}) 
and on Oct.\,23 at the east limb 5$^o$N, 15$^o$S, 23$^o$S and 32$^o$S ({\it 
lower left panel}) together with the corresponding disk appearance on Oct.\,25  
({\it lower right panel}).}
   \label{Fig1}
    \end{figure}

%
%
\section{Results for the H$\alpha$ and H$\beta$ lines}

\subsection{Emission relations}

The observed emissions, given in Fig.\,3, reach line a radiance up to 
$E\beta = 16\cdot 10^4$ erg/(s cm$^2$ ster), being four times larger 
than the maximum values by Stellmacher \& Wiehr (1994b; their Figs.\,2 
and 3). Comparably high values were published by Yakovkin \& Zel'dina (1963; 
entered in Fig.\,3). For faint $E\beta\le 1\cdot 10^4$ erg/(s cm$^2$ ster), 
our data perfectly match the general empirical relation by Stellmacher \& Wiehr 
(1994b) between E$\alpha$ and E$\beta$ (crosses in Fig.\,3). 
For stronger emissions  $E\beta > 3\cdot 10^4$ our recent data are slightly below 
that curve. The first Balmer decrement E$\alpha$/E$\beta$ shows for the brightest 
emissions values near 3.0, and limiting values clearly above 10.0 for the faintest 
emissions. This agrees with Stellmacher \& Wiehr (1994b) and follows the curves 
(also entered in Fig.\,3) calculated by Gouttebroze et al. (1993; hereafter referred 
to as GHV) for temperatures of $T_{kin} = 4300$\, K and, respectively, $T_{kin} = 6000$\,K. 

The here obtained values of line radiance well agree with former data from much fainter 
and more structured prominences, obtained with quite different methods. This indicates 
that possible influences from 'filling' play a minor role (in agreement with Stellmacher, 
Wiehr, Dammasch 2003), as is also seen from the independence on the aspect angle: e.g., 
we do not find significant differences between the 'face-on' prominence at $5^o$N and the
'end-on' one at $32^o$S, both at the east limb on Oct. 23 (marked in Fig.\,1).
                    
%

   \begin{figure}[ht]     
   \includegraphics[width=8.8cm]{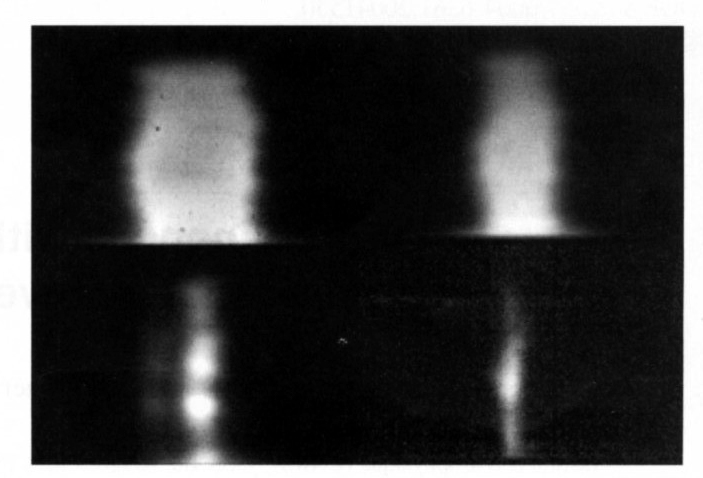}
   \caption{CCD spectra of the Balmer lines H$\alpha$ ({\it upper left}), 
H$\beta$ ({\it upper right}), both with central reversions, and the 
simultaneously observed  HeD$_3$ ({\it lower left}) and  NaD$_2$
({\it lower right panel}), in the prominence at W/24$^o$S on Oct. 18; 
each sub-image spans 41" x 2.5\AA.}
   \label{Fig2}
    \end{figure}

%
%
                    
%

   \begin{figure}[htb]     
   \hspace{0mm}\includegraphics[width=8.5cm]{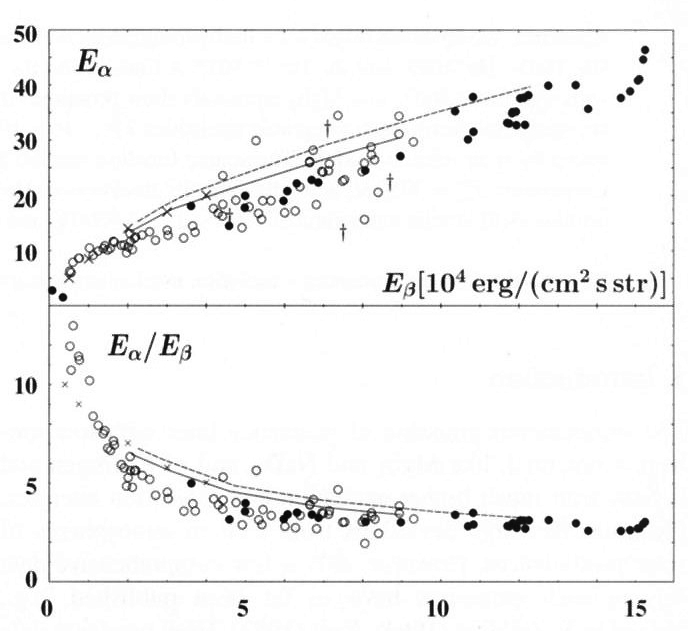}
   \caption{Observed line radiance E$\alpha$ ({\it upper}) and 1.\,Balmer 
decrement E$\alpha$/E$\beta$ ({\it lower panel}) both versus E$\beta$; 
prominences with strong central H$\alpha$ reversion ({\it filled circles}); 
mean observations by Stellmacher \& Wiehr (1994b) ({\it crosses}); data by
Yakovkin \& Zel'dina (1963) ({\it dags}); model calculations by GHV (1993) 
for $T_{kin} = 4300$\,K ({\it solid}) and $T_{kin} = 6000$\,K ({\it dashed line}).}
    \label{Fig3}
    \end{figure}

%
%

\subsection{Centrally reversed H$\alpha$ profiles}

We observe distinct central reversions (double peaks) of the  H$\alpha$ profile 
for a line radiance $E\beta > 5\cdot 10^4$ erg/(s\,cm$^2$\,ster). An example is 
shown in Fig.\,4 together with simultaneously observed He and metal lines. 
We find the most prominent central reversions in the strongest, yet {\it narrow} 
emission profiles. In Fig.\,5 we give some relations of the observed reversion 
signatures and compare them to such from the comprehensive set of H$\alpha$ emission 
profiles calculated by GHV for thick slabs of models with $T_{kin} = 6000-8000$\,K 
and $v_{nth} = 5$\,km/s, acting as semi-infinite layers. 

Profiles with the most prominent central reversions are markedly narrow, and 
their signatures (filled circles in Fig.\,5) well follow the calculated 
relations and can thus be explained by pure line-saturation. Data that deviate 
from the calculated curves (open circles) are derived from broader line profiles. 
This stronger broadening (e.g., due to macro velocities or superpositions) will 
readily lead to a deterioration of the pure saturation effect.

The fact that the bright and rather unstructured prominences show strikingly narrow 
lines was already mentioned by Stellmacher \& Wiehr (1994b). We consider the central 
reversions as a signature of emission in semi-infinite dense layers. No evident 
relation is found between the intensity difference of the two emission peaks near 
line-center. Here, non-LTE transfer calculations should be considered, including 
'spatially correlated velocity fields' as suggested by Magnan (1976).

%

   \begin{figure}[htb]     
   \hspace{-3mm}\includegraphics[width=9.5cm]{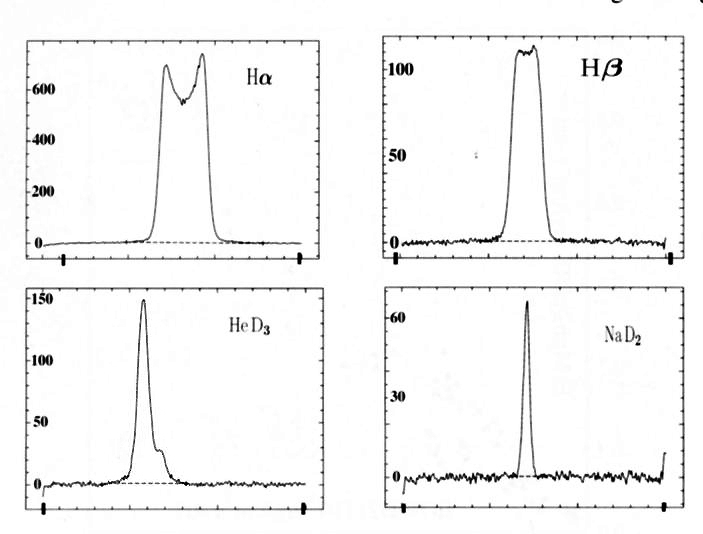}
   \caption{Profiles of simultaneously observed emissions with distinct saturation of 
 H$\alpha$ and H$\beta$ ({\it upper panels}) together with HeD$_3$ and NaD$_2$ 
({\it lower levels}): ordinates = CCD counts; thick abscissa tick marks give
$\Delta\lambda = 5.0$\AA{}; (He-CCD misc-entered).}
    \label{Fig4}
    \end{figure}
                  
%

   \begin{figure}[htb]     
   \hspace{0mm}\includegraphics[width=8.5cm]{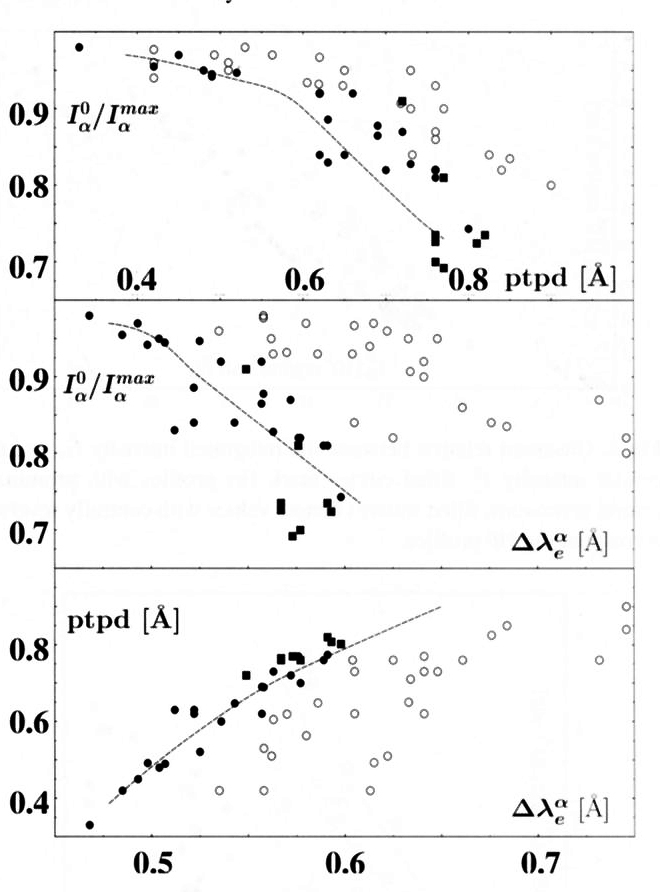}
   \caption{Observed relations of central intensity $I^0_{\alpha}$, peak-intensity 
$I^{max}_{\alpha}$, peak-to-peak distance ptpd of the two H$\alpha$ maxima, and the half 
line width $\Delta\lambda_e$ at $I^0_{\alpha}/e$; emissions with prominent central 
reversals are marked by filled circles (as in Fig. 3); filled squares denote values 
for which corresponding H$\beta$ profile is also centrally reversed. {\it Dashed 
lines} give  calculations from slab-models by GHV.}
    \label{Fig5}
    \end{figure}

%
%

\subsection{Central intensities and source function}

The double peaked H$\alpha$ emission originates from thick layers with
$\tau_0^{\alpha}>>1.0$, for which the central line intensity becomes 
$I_0^{\alpha} = S^{\alpha}(1 - e^{-\tau_0^{\alpha}})\approx S^{\alpha}$ 
allowing us to directly deduce the source function. For the strongest emissions, 
we find, (following the method described by Stellmacher \& Wiehr 1994b; Sect.\,4) 
$\tau_0^{\alpha}>>5.0$. If we express the relative level population of the 
H$\alpha$ transition by the Boltzmann formula: 
$$(n_{0,3}g_{0,3})/(n_{0,2}g_{0,2})=exp[-hc/(\lambda_{\alpha}k\,T^{\alpha}_{ex}]$$
and insert this into the general equation for the source function, we obtain the 
corresponding Planck function, $B$, for the excitation temperature $T^{\alpha}_{ex}$:
 
$$ S_{\alpha}=(2\,h\,c^2/\lambda_{\alpha}^5)/ (exp[-hc/(\lambda_{\alpha}k\,T^{\alpha}_{ex})]-1)=
B(T^{\alpha}_{ex})$$

The mean upper values (open circles in Fig.\,6) for non-reversed profiles with 
$I_0^{\alpha} \approx 36\cdot 10^4$ erg/(s\,cm$^2$\,ster\,\AA{}) correspond to an excitation 
temperature $T^{\alpha}_{ex} \approx 3950$\,K. Fainter prominences analyzed by Stellmacher 
\& Wiehr (1994b) gave smaller mean upper values $I_0^{\alpha} \approx 26\cdot 10^4$
corresponding to $T^{\alpha}_{ex} \approx 3700$\,K. GHV obtain for their models with 
$T_{kin} = 6000$\,K and $v_{nth} = 5$ km/s a source function $S^{\alpha}=36\cdot 10^4$ at 
$\tau_0^{\alpha} = 10$ in good agreement with the present observations (Fig.\,6).
 
The central intensity of the reversed profiles reaches values  
$I_{max}^{\alpha}\approx 43\cdot 10^4$ erg/(s\,cm$_2$\,ster\,\AA{}) (cf., Fig.\,5) 
which corresponds to $T^{\alpha}_{ex}\approx 4000$\,K. The two peaks outside the line 
center arise from smaller $\tau^{\alpha}$ values, their higher intensity then indicating 
an increase of the source function towards the prominence interior (see also Yakovkin 
\& Zel'dina 1964). The model calculations of GHV (1994; their Fig.\,18) indicate 
a rise of $S^{\alpha}$ even beyond $0.16\cdot I^{phot}$, i.e. $S_{\alpha} > 46\cdot 10^4$ 
erg/(s\,cm$_2$\,ster\,\AA{}); this conflicts with the occurrence of dark filaments 
on the disk. Such high $S^{\alpha}$ might only be valid for spicules and 
eruptive objects.
                    
%

   \begin{figure}[htb]     
   \hspace{2mm}\includegraphics[width=9.0cm]{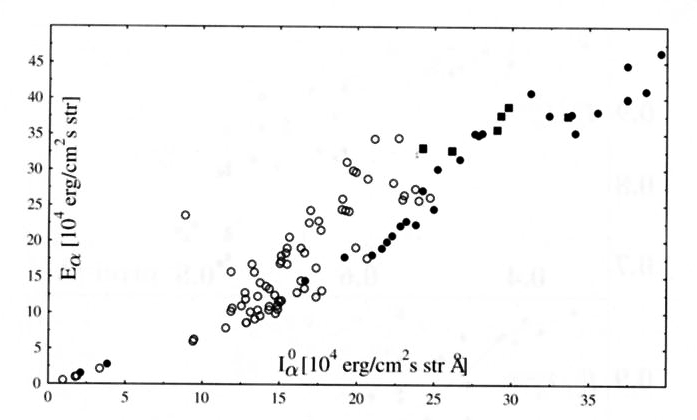}
   \caption{Observed relation between the integrated intensity $E\alpha$ and the central 
intensity $I^0_{\alpha}$; filled circles mark H$\alpha$ profiles with prominent central 
reversions, filled squares denote values for which the corresponding  H$\beta$ profile 
is also centrally reversed.}
    \label{Fig6}
    \end{figure}

%
%
                    
%

   \begin{figure}[htb]     
   \hspace{2mm}\includegraphics[width=9.0cm]{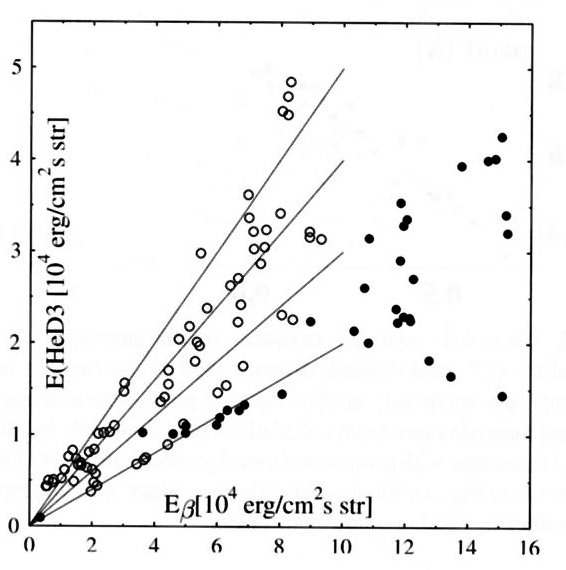}
   \caption{Integrated intensities of HeD$_3$ and H$\beta$; emissions corresponding to 
prominent H$\alpha$ reversions are marked by filled circles; full lines trace emission 
ratios of 0.5, 0.4, 0.3, and 0.2}
    \label{Fig7}
    \end{figure}

%
%
                    
%

   \begin{figure}[htb]     
   \hspace{0mm}\includegraphics[width=9.cm]{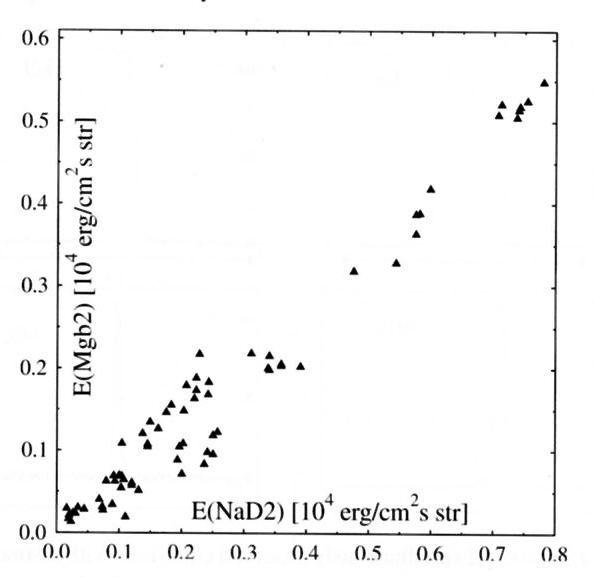}
   \caption{Emission relation of Mgb$_2$ and NaD$_2$.}
    \label{Fig8}
    \end{figure}

%
%

\section{The helium emissions}

The simultaneously observed HeD$_3$ and H$\beta$ lines show distinct branches in their 
intensity relation (Fig.\,7): Emissions from prominence locations with prominent H$\alpha$ 
reversions show ratios E$_{HeD3}/E_{\beta} = 0.2 - 0.3$, while less thick prominences 
follow branches with ratios 0.4 - 0.5. This confirms earlier results by Stellmacher 
\& Wiehr (1994a, 1995) who found from analyses of He\,3889 with H\,3888. and of HeD$_3$ 
with H$\beta$ that the emission ratio of the He-triplet and the Balmer lines shows 
typical mean values, in the sense that prominences with stronger Balmer emissions 
(known to be cooler, less structured, and denser; cf., introduction) yield lowest 
Helium-to-Balmer ratios.

The HeD$_3$ fine-structure components can be used to measure of the optical thickness, as
was shown by Stellmacher, Wiehr, Dammasch (2003) for the analogous case of He\,10830\AA{}. 
In contrast to that triplet line, HeD$_3$ does not allow us to determine the ratio of the 
faint red component and the two (not separated) main components with similar reliability, 
due to the much smaller spectral distance of only 0.32\,\AA{}. The ratio  1\,:\,8 for the 
optically thin case increases up to 1\,:\,6 for emissions 
E$_{HeD3}\ge 0.6\cdot 10^4$ erg/(s\,cm$^2$\,ster), indicating that HeD$_3$ begins to saturate.

Our instrumental set-up also covers the lines He\,II\,4685.7 and He\,I\,5015.7 (singlet). 
We do not find any significant He\,II emission in the prominences observed, which may be 
too cool and dense for sufficient He\,II excitation. The faint He\,I singlet line was 
only measurable in one prominence, its width is quite similar to that of HeD$_3$, yielding 
$(\Delta\lambda/\lambda_0)^{He\,I}\le 2.7\cdot 10^{-5}$. The integrated 
intensity E(He5016)$\approx 0.05\cdot 10^4$ erg/s\,cm$^2$\,ster gives an emission 
ratio with HeD$_3$ of He$^{singl}$/He$^{tripl}\approx 0.016$. Other line combinations, 
including the stronger singlet line He\,6678, might be useful to extend the study of 
singlet-to-triplet ratio by Stellmacher \& Wiehr (1997).
                    
%

   \begin{figure}[htb]     
   \hspace{-1mm}\includegraphics[width=9.2cm]{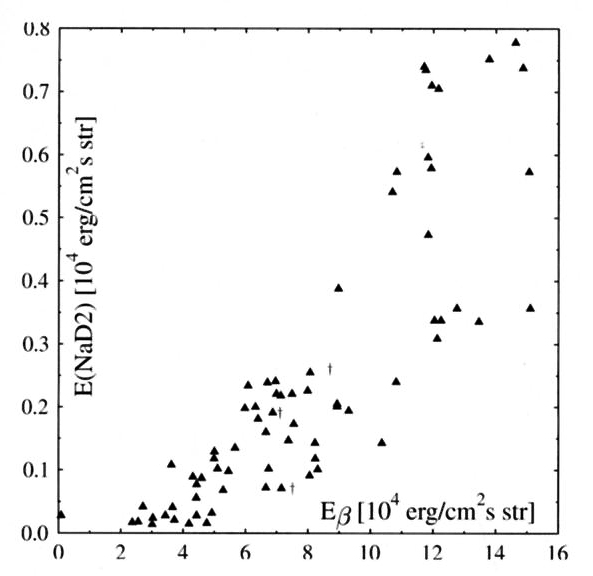}
   \caption{Integrated intensity of NaD$_2$ versus that of H$\beta$; 
dags show the observations by Yakovkin \& Zel'dina ( 1963).}
    \label{Fig9}
    \end{figure}

%
%

\section{Emissions of the metal lines Mgb$_2$ and NaD$_2$}                    

The integrated intensities of Mgb$_2$ and NaD$_2$ show a linear relation with a ratio 
E(Mgb$_2$)/E(NaD$_2$)= 0.7 (Fig.\,8), indicating that the emissions of both lines are 
closely related. Their integrated intensities strongly depend on the prominence 
thickness, as can be seen from the relation with H$\beta$ in Fig.\,9. We observe 
reliable emissions of NaD$_2$ only if E$_{\beta}>3\cdot 10^4$ erg/s\,cm$^2$\,ster. 
Maximum E(NaD$_2) \approx 0.78\cdot 10^4$ erg/s\,cm$_2$\,ster is observed at 
prominence locations with  E$\beta > 10\cdot 10^4$ erg/s\,cm$^2$\,ster, where 
the H$\beta$ profiles are saturated or even centrally reversed.
  
The mean Doppler widths of the observed metal lines amount to 
$\Delta\lambda_D^{NaD_2} \approx 95$ and $\Delta\lambda_D^{Mgb_2} \approx 82$\,m\AA{},
corresponding to $\Delta\lambda_D/\lambda \approx 1.6\cdot 10^{-5}$. The narrow widths 
from this almost entirely non-thermal broadening yield $v_{nth}\approx 5$\,km/s. 
Similarly narrow profiles of Mgb$_2$ were reported by Landman (1985). Comparison with 
model calculations by Kim (1987; Fig.\,7) indicates that these observations can only 
be reproduced with high total number densities up to $N\le 10^{12}$\,cm$^{-3}$.

\section{Conclusions}

The present spectro-photometry extends our former analyses (Stellmacher \& Wiehr 1994b, 
1995) to four times higher H$\beta$ emissions. The here observed bright prominences are 
low latitude objects at $\vartheta < 32^o$. We find prominent central reversions of 
H$\alpha$ and occasionally of H$\beta$ for line radiance E$\beta> 5\cdot 10^4$ 
erg/(s\,cm$^2$\,ster). These centrally reversed profiles can well be modeled assuming 
semi-infinite layers, as is seen from a comparison with model calculations by Gouttebroze 
et al. (1993). The emitting layers should then consist of 'densely wound fibers forming 
massive ropes or wicks' (cf., Engvold 1998).

THEMIS proved to be a powerful instrument for multi-line spectral photometry, also useful 
for solar prominences. Due to its low stray-light level (seen in the rather faint aureole 
spectra) and its low instrumental polarization, one may extend these observations to 
filtergram techniques (cf., Stellmacher \& Wiehr 2000) for a study of the dynamics of 
small-scale prominence structures inclusive their magnetic field as, e.g., done by 
Wiehr \& Bianda (2003).

%
%

\begin{acknowledgements}
We thank the THEMIS team, in particular C.\,Briand for kind support. We are indebted 
to ENO for the grant 'THEMIS project No.\,42'. The THEMIS telescope on Tenerife is operated 
by the French 'Centre National de la Recherche Scientifique' and the Italian 'Consiglio 
Nazionale delle Ricerche' at the Spanish 'Observatorio del Teide' of the Instituto de 
Astrof\'isica de Canarias.
\end{acknowledgements}

\eject
%
%

%
%


\begin{thebibliography}{}

\bibitem{} Engvold, O. 1998, in (D.\,F. Webb, B. Schmieder, D.\,M. Rust, eds.) 
Proc. 'New Perspectives on Solar Prominences', IAU-coll. 167, 23

\bibitem{} Gouttebroze, P., Heinzel, P. \& Vial, J.\,C. 1993, A\&AS 99, 513 

\bibitem{} Gouttebroze, P., Heinzel, P. \& Vial, J.\,C. 1994, A\&A 292, 656 

\bibitem{} Kim, K.\,S. 1987, Solar Phys. 114, 47 

\bibitem{} Labs, D. \& Neckel, H. 1970, Solar Phys. 15, 79 

\bibitem{} Landman, D.\,A. 1985, ApJ 295, 220

\bibitem{} Magnan, C. 1976, J. Quant. Spectrosc. Radial. Transfer 16, 281 

\bibitem{} Stellmacher, G. \& Wiehr. E. 1994a, A\&A 286, 302 

\bibitem{} Stellmacher, G. \& Wiehr, E. 1994b, A\&A 290, 665 

\bibitem{} Stellmacher, G. \& Wiehr, E. 1995, A\&A 299, 921 

\bibitem{} Stellmacher, G. \& Wiehr, E. 1997, A\&A 319, 669 

\bibitem{} Stellmacher, G. \& Wiehr, E. 2000, Solar Phys. 196, 357 

\bibitem{} Stellmacher, G., Wiehr, E. \& Dammasch, I.\,E. 2003, Solar Phys. 217, 133

\bibitem{} Wiehr, E. \& Bianda, M. 2003, A\&A 404, L25

\bibitem{} Yakovkin, N. A. \& Zel'dina, M.\,Yu. 1963, Astron. Zh. 40, 847

\bibitem{} Yakovkin, N. A. \& Zel'dina, M.\,Yu. 1964, Astron. Zh. 41, 914

\end{thebibliography}
\end{document}